\documentclass[twocolumn, notitlepage, secnumarabic,amssymb, aps, prl, figure, superscriptaddress, floatfix] {revtex4-1}
\usepackage{graphicx}
\usepackage{amsmath}
\usepackage{subfigure}
\usepackage{braket}
\usepackage{ulem}
\usepackage{hyperref}
\usepackage{bm}
\hypersetup{
    colorlinks=true,
    linkcolor=blue,
    filecolor=red,      
    urlcolor=blue,
    citecolor=magenta
}
\usepackage[version=4]{mhchem}
\usepackage{ragged2e}
\usepackage{float}

\usepackage[usenames, dvipsnames]{color}
\definecolor{mypink}{RGB}{219, 48, 122}

\begin{document}

\title{A Sparse Bayesian Committee Machine Potential for Hydrocarbons}

\author{Soohaeng Yoo Willow}
\affiliation{Department of Energy Science, Sungkyunkwan University, Seobu-ro 2066, Suwon, 16419, Korea}

\author{Gyung Su Kim}
\affiliation{Department of Energy Science, Sungkyunkwan University, Seobu-ro 2066, Suwon, 16419, Korea}

\author{Miran Ha}
\affiliation{Department of Chemistry, Ulsan National Institute of Science and Technology, 50 UNIST-gil, Ulsan 44919, Korea}

\author{Amir Hajibabaei}
\affiliation{Yusuf Hamied Department of Chemistry, University of Cambridge, Lensﬁeld Road, Cambridge, CB2 1EW, United Kingdom}

\author{Chang Woo Myung}
\email{cwmyung@skku.edu}
\affiliation{Department of Energy Science, Sungkyunkwan University, Seobu-ro 2066, Suwon, 16419, Korea}

\date{\today}

\begin{abstract}
Accurate and scalable universal interatomic potentials are key for understanding material properties at the atomic level, a task often hindered by the steep computational scaling. Although recent developments of machine learning potential has made significant progress, the flexibility and expansion to a wide range of compounds within a single model seems still challenging to build in particular for the kernel-based models. Here, we introduce the Bayesian Committee Machine (BCM) potential to handle a wide array of hydrocarbons in various phases (gas, clusters, liquid, and solid). The BCM potential leverages the committee approach to bypass the poor scaling of kernel regressors when dealing with large datasets. Its committee-based structure allows for easy and cost-effective expansion, maintaining both transferrability and scalability. Demonstrating its robustness, the BCM accurately models challenging examples like the Diels-Alder reaction, structural strains, and pi-pi interactions. Our systematic benchmarking positions the sparse BCM potential as a promising candidate in the quest for a universal ab initio machine learning potential. \end{abstract}

\maketitle

\section{Introduction}

\begin{figure*} [htp]
\includegraphics[width=6.in]{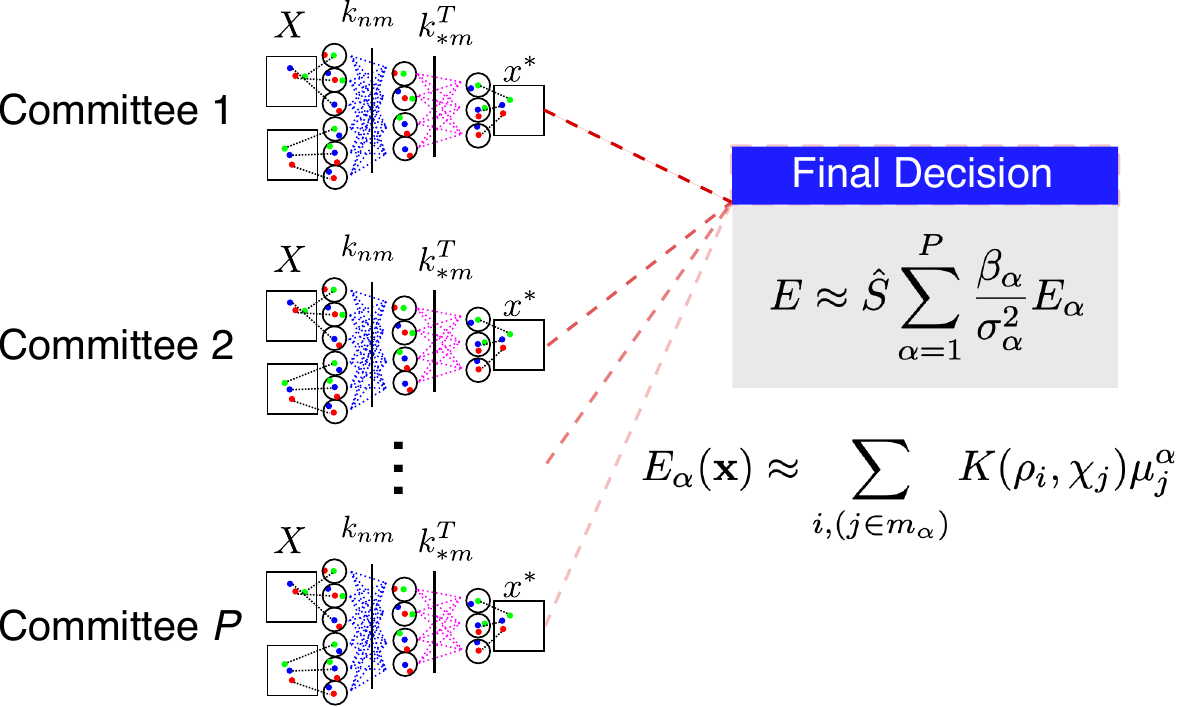}
\caption{Schematic of Bayesian committee machine potential where $P$ committee models (sparse local experts in this work) make a final decision on any properties of system weighted by $\frac{\beta_\alpha}{\sigma^2_\alpha}$. The weight ensures to have confident predictions to be more outspoken than less confident models. Different forms of weights $\beta_\alpha$ of $\alpha$-th expert give rise to BCM ($\beta_\alpha=\beta_0$) and RBCM ($\beta_\alpha=log(\sigma_{**}^2)-log(\sigma_{i}^2)$). 
}
\label{fig:schematic}
\end{figure*}

Gaussian processes (GPs) are a fundamental tool in statistical non-parametric learning used for tasks like regression, classification, sampling, multi-task learning, and more.\cite{rasmussenGaussianProcessesMachine2005,shahriariTakingHumanOut2016,lawrenceProbabilisticNonlinearPrincipal2005,alvarezComputationallyEfficientConvolved2011,NIPS2000_19de10ad,liuWhenGaussianProcess2019,gaoGeneralizedLocalAggregation2020,deisenroth2015distributed,liu2018generalized,trespBayesianCommitteeMachine2000}
When trained on a dataset consisting of input vectors $X = \{x_i \in \mathbb{R}^d\}_{i=1}^n$ and corresponding observations $y = \{y(x_i) \in \mathbb{R}\}_{i=1}^n$, where $n$ is the size of the training data set, $\mathbb{R}^d$ represents the $d$-dimensional vector space of real numbers, GPs make accurate predictive distributions for the latent function $\eta : \mathbb{R}^d \rightarrow \mathbb{R}$. As the demand for data-driven methodologies in scientific and engineering fields increased, especially in materials science, using an ab initio level machine learning potential for atomistic simulations has emerged as a vital tool to accelerate accurate atomistic simulations. These models must be capable of handling complex and high-dimensional potential energy surface representing atomic interactions. Ab initio machine learning potentials, which integrate first-principle quantum mechanics with data-driven machine learning, demand precise and scalable statistical models to capture intricate and high-dimensional ab initio interatomic interactions. Among various ML models,\cite{behlerGeneralizedNeuralNetworkRepresentation2007,behlerAtomcenteredSymmetryFunctions2011,zhangDeepPotentialMolecular2018,zhangAdaptiveCouplingDeep2018,schranMachineLearningPotentials2021,atzGeometricDeepLearning2021,reiserGraphNeuralNetworks2022,eckhoffHighdimensionalNeuralNetwork2021,schütt2021equivariant} including neural networks,\cite{behlerGeneralizedNeuralNetworkRepresentation2007,behlerAtomcenteredSymmetryFunctions2011,zhangDeepPotentialMolecular2018,zhangAdaptiveCouplingDeep2018,schranMachineLearningPotentials2021} graph networks,\cite{atzGeometricDeepLearning2021,reiserGraphNeuralNetworks2022} message-passing,\cite{schütt2021equivariant} the GP-based ML potential boasts in the active learning, uncertainty prediction, and low data requirement. However, despite all these advantages, the scalability of GP-based ML potential\cite{trespBayesianCommitteeMachine2000,deisenroth2015distributed,liu2018generalized} came under scrutiny in particular for achieving universal potential due to its poor scaling. 

In the standard setup, ML potentials based on GPs face significant computational challenges with large training sets. The complexity of training a GP increases dramatically as as $O(n^3)$ with the size of the training set $n$  because it involves inverting a Kernel matrix. When it comes to making predictions, the process becomes quadratic with the size of the training set. Due to these computational hurdles, GPs struggle to handle datasets larger than the order of $10^4$. The challenge of handling large datasets with GPs led to the development of sparsification techniques. To balance the need for precision with computational efficiency, the authors introduced sparse Gaussian process regression (SGPR) potential that uses a chosen subset of the data (inducing points) to represent the full GP potential.\cite{quinonero-candelaUnifyingViewSparse2005,hajibabaeiSparseGaussianProcess2021,hajibabaeiMachineLearningFirstPrinciples2021,hajibabaeiUniversalMachineLearning2021,haSparseGaussianProcess2022,myungChallengesOpportunitiesProspects2022}
Although this reduces computational demands, it still faces challenges when the number of inducing points $m$ is large, 
since the computational demands scale at $O(nm^2)$. 

In response to this, attention turned towards aggregation models, specifically the Bayesian Committee Machine (BCM).\cite{deisenroth2015distributed,liu2018generalized,trespBayesianCommitteeMachine2000,hintonTrainingProductsExperts2002,caoGeneralizedProductExperts2015}
These models combine predictions from multiple sub models, or GP experts, each trained on different sections of the data space. We argue that this approach is particularly useful for building ab initio machine learning potential for wide range of materials. BCM utilizes Bayesian approach to scale up GP learning, making it suitable for combining complex interatomic interactions. Aggregation type models like BCM allows efficient distribution of model tasks and better opportunities for parallel processing of GP models. However, it was pointed out a particular issue with common aggregation methods  product-of-experts (PoE)\cite{hintonTrainingProductsExperts2002,caoGeneralizedProductExperts2015} 
and Bayesian committee machine (BCM)\cite{trespBayesianCommitteeMachine2000,liu2018generalized,deisenroth2015distributed}
that in certain types of training data, these methods do not  provide consistent predictions as the training data size grow large $n \to \infty$.\cite{rulliereNestedKrigingPredictions2018}
To further improve this inconsistency, various committee models, such as the nested pointwise aggregation of experts (NPAE), robust BCM (RBCM), the generalized RBCM (GRBCM), were develop by removing the conditional independence assumption condition. \cite{gaoGeneralizedLocalAggregation2020}

In this work, by leveraging powerful scaling strategies of GPs - sparsification and aggregation - we introduce the BCM potential that is highly scalable and accurate and apply this to the example of wide class of hydrocarbons. We tested the BCM potential by comparing the energy, forces, vibrational frequencies of various hydrocarbon classes. We also demonstrate the robustness of the BCM potential by comparing the reaction pathways, steric effects, $\pi$-$\pi$ interactions against the density functional theory calculations. This forges a way to achieve highly scalable GP-based ML potential towards universal potential. 

\section{Theory\label{sec:theory}}

\begin{figure*} [htp]
\includegraphics[width=17cm]{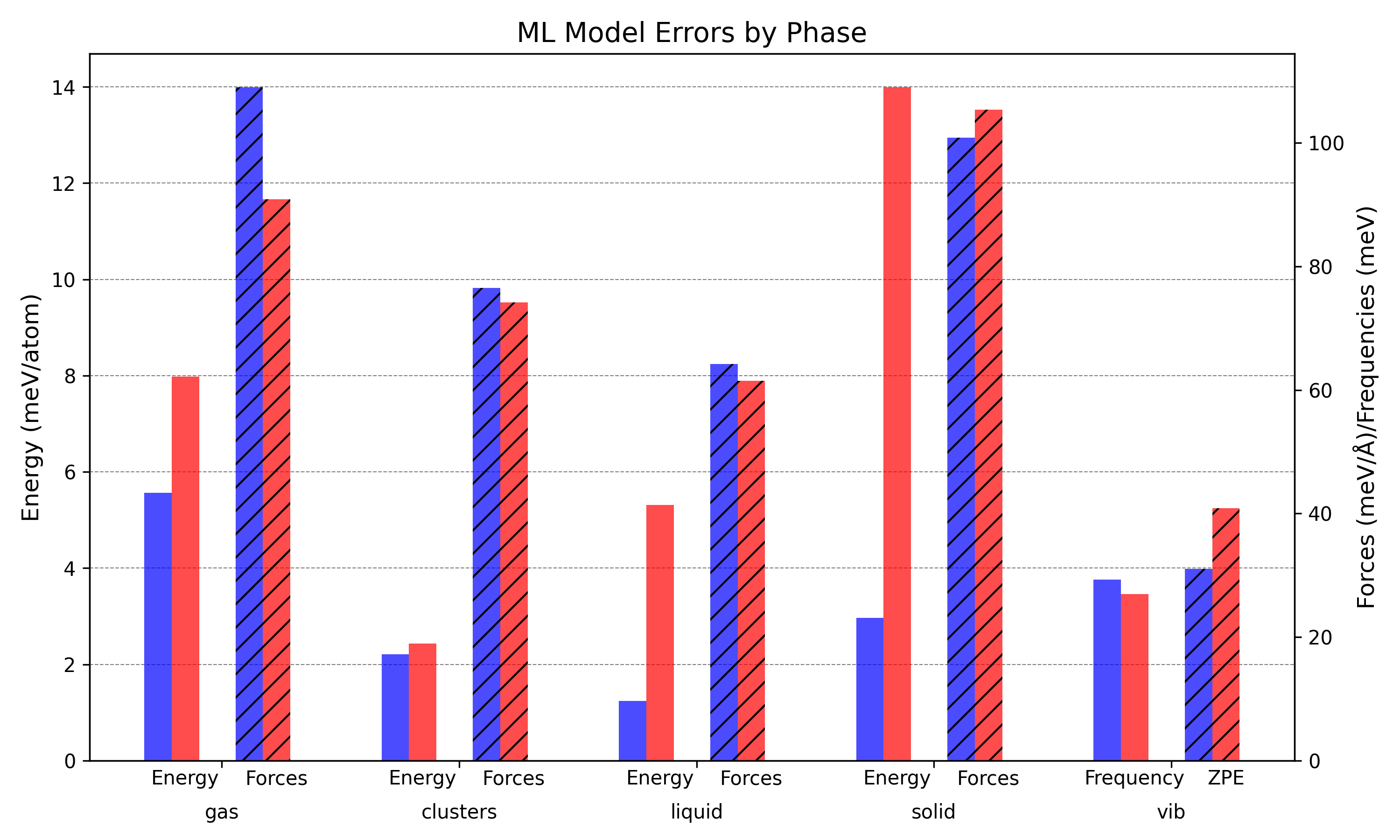}
\caption{Overall test mean absolute errors (MAEs) across different phases  of ML potential models, sparse local experts (SLE) and robust Bayesian committee machine (RBCM). MAEs of test sets for energy (meV/atom) and forces(meV/Å) are calculated for gas phase, molecular clusters, liquid, and solid systems, as well as vibration frequencies and zero-point energy (ZPE). SLE and RBCM models are depicted in blue and red bars, respectively.}
\label{fig:test_errors}
\end{figure*}

A configuration of $N$ atoms, denoted as $x$, is transformed into a list of descriptors $\bm{x} = \{\rho_i\}_{i=1}^{N}$, where $\rho_i$ represents a rotationally invariant descriptor characterizing the local chemical environment (LCE) of atom $i$. This descriptor is solely determined by the relative coordinates of the $N-1$ atoms within a specified cut-off radius.
In SGPR,\cite{quinonero-candelaUnifyingViewSparse2005,hajibabaeiSparseGaussianProcess2021}
the potential energy is defined as
\begin{eqnarray}
E(\bm{x}) & = & \sum_{i=1}^N\sum_{j=1}^{m} K(\rho_i, \chi_j) \mu_j.
\end{eqnarray}
Additionally, the forces can be expressed as:
\begin{eqnarray}
F_{i} & = & -\sum_{j=1}^{m} K(\dot{\rho_i}, \chi_j) \mu_j.
\end{eqnarray}
where $\dot{\rho_i}$ represents the gradient of the atomic density in the neighborhood of atom $i$ with $\bm{\dot{x}} = \{\dot{\rho}_i \}_{i=1}^N$.
Here, $\bm{z} = \{\chi_j\}_{j=1}^m$ represents the set of inducing descriptors, 
$K$ is a covariance kernel, 
and $\bm{\mu} = \{\mu_j\}_{j=1}^m$ represents the vector of weights associated with the inducing descriptors. 
The weight vector $\bm{\mu}$ is determined in such a way that 
it reproduces potential energy, forces, and virial stress $\bm{Y}$ for a given dataset 
$\bm{X} = \{\rho_k, \dot{\rho}_k\}_{k=1}^n$ as the solution to the following SGPR equation:
\begin{eqnarray}
\begin{bmatrix}
\bm{k}_{nm} \\
\sigma \bm{L}^{T}
\end{bmatrix}
\bm{\mu} = 
\begin{bmatrix}
\bm{Y} \\
\bm{0}
\end{bmatrix},
\end{eqnarray}
where $\bm{k}_{nm}$ represents the data-inducing ($\bm{X}-\bm{z}$) covariance matrix, 
$\sigma$ is the noise hyperparameter, and 
$\bm{L}$ is the Cholesky factor of $\bm{k}_{mm}$, which represents the covariance matrix between inducing descriptors $\bm{z}$.
The inducing descriptors are a subset of descriptors extracted from $\bm{X}$. 

We build the BCM potential by actively sampling the LCE whose covariance loss $s(\rho_i)$ is larger than a predefined threshold. The covariance loss is defined in 
\begin{eqnarray}
    s(\rho_i) = K(\rho_i, \rho_i) - k_{\rho_i m} k^{-1}_{mm} k_{\rho_i m}^{T}.
\end{eqnarray}

When both the dataset $\bm{X}$ and the number of inducing descriptors increase, it demands a substantial amount of memory and CPU resources, $\mathcal{O}(m^2)$.
To enhance memory efficiency and increase the parallelization of the SGPR algorithm, we partition
the dataset and inducing descriptors into $p$ distinct subsets: 
 $m_1, \cdots, m_p$. 
Leveraging the BCM algorithm,
we approximate the predicted potential energy as
\begin{eqnarray} \label{eq:E_bcm}
    E \approx \hat{S} \sum_{\alpha=1}^{p} \frac{\beta_\alpha}{\sigma_\alpha^2}E_\alpha,
\end{eqnarray}
where 
the potential energy $E_\alpha$ is determined by utilizing
inducing descriptors $\bm{z}_\alpha = \{\chi_j\}_{j \in m_\alpha}$, along with the dataset $\bm{X}_\alpha = \{\rho_k, {\dot{\rho}}_k \}_{k \in m_\alpha}$,
and the weight vector $\bm{\mu}^\alpha$ of $\alpha$-th sub SGPR model.
This energy can be represented as:
\begin{eqnarray} \label{eq:E_alpha}
    E_\alpha (\bm{x}) & \approx & 
    \sum_{i=1}^N \sum_{j \in m_\alpha}  K(\rho_i, \chi_j) \mu_j^\alpha. 
\end{eqnarray}
The maximum value of the covariance loss for the $\alpha$-th dataset and inducing descriptors, denoted as $\sigma_\alpha^2$, is used to weight the $\alpha$-th committee prediction. Furthermore, we weight the prediction with $\beta_\alpha$ following the concepts of robust BCM where the individual committee prediction is weighted by the differential entropy, $\beta_\alpha = \log(\sigma_{**}^2)-\log(\sigma_{i}^2)$.\cite{caoGeneralizedProductExperts2015,liu2018generalized} The final normalization factor $\hat{S}$ is thus given:
\begin{eqnarray}
 \hat{S} &= &\left[ \sum_\alpha^p \frac{\beta_\alpha}{\sigma_\alpha^2}\right]^{-1}
\end{eqnarray}

Our previous attempt was a half measure where the SGPR model is only built locally to a specific subset of hydrocarbons.\cite{haSparseGaussianProcess2022} However, to build a transferable ML potential for a wide range of hydrocarbons, one needs a unified single entity model.\cite{hajibabaeiUniversalMachineLearning2021} In this work, we employed the robust BCM (RBCM), represented by $\beta_\alpha = \log(\sigma_{**}^2)-\log(\sigma_{i}^2)$. 
Note that the BCM potential energy is close to a potential energy of a particular local expert when the weight $\beta_\alpha$ is defined as $\beta_\alpha = \sigma_\alpha^{-\infty}$.


\begin{figure*} [htp]
\includegraphics[width=17cm]{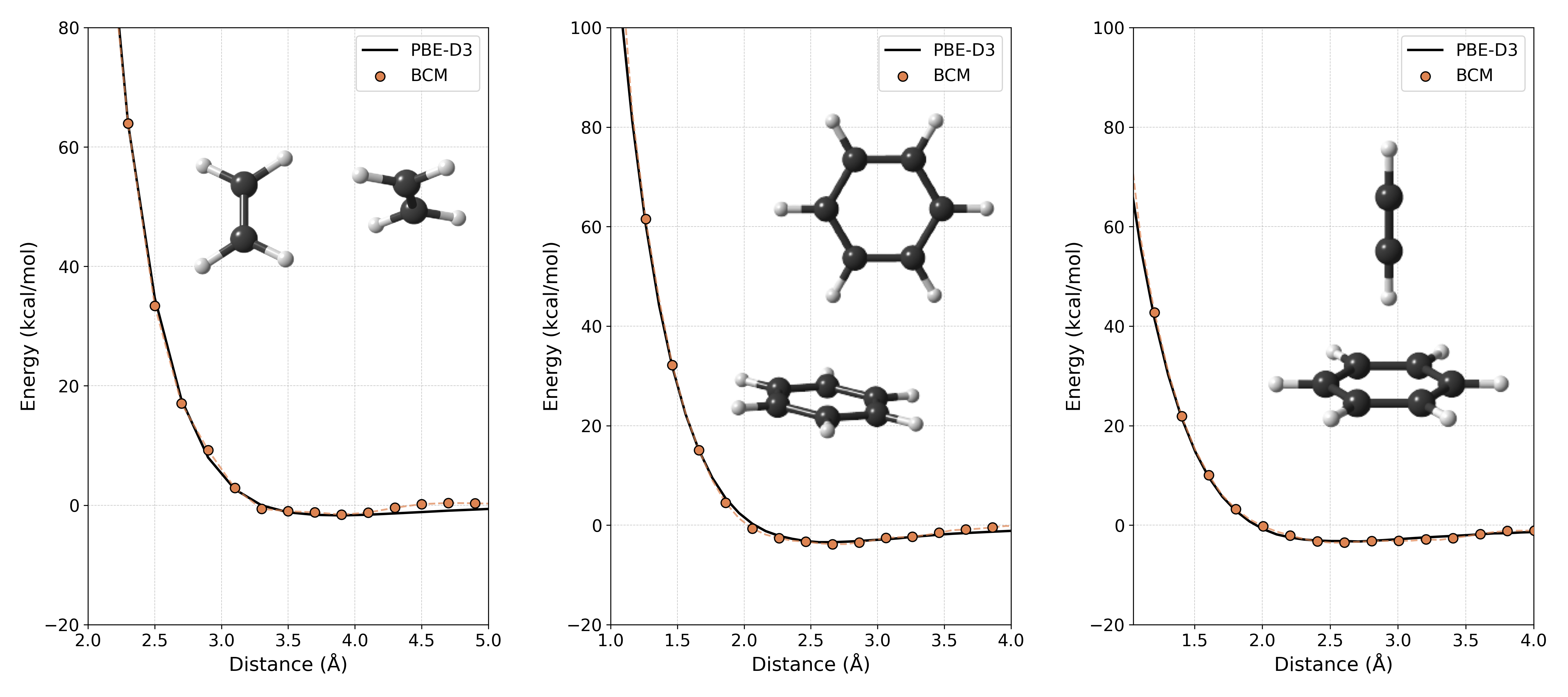}
\caption{Comparative dissociation energy curves of $\pi$-$\pi$ dimer hydrocarbons. (a) Dissociation curve (kcal/mol) of edge-on-face ethylene dimer with the intermolecular distance (Angstrom) defined between the centroids of the ethylene molecules. (b) Dissociation curve (kcal/mol) of benzene dimer adopting a tilted T-shaped orientation. The distance is defined between the centroid of one benzene molecule and the nearest hydrogen of its counterpart. (c) Dissociation curve (kcal/mol) of T-shaped benzene-ethyne dimer with the distance between the benzene centroid and the proximate hydrogen of ethyne.}
\label{fig:pi-pi}
\end{figure*}

\section{Results}

We built local sub-expert SGPR potentials, denoted as $E_\alpha$ in Eq. (\ref{eq:E_alpha}), for different hydrocarbons.
These hydrocarbons encompass alkane, alkene, isoalkane, alkadiene, alkatriene, cycloalkane, cycloalkane-methyl, cycloalkene, bicycloalkane, aromatic, and polyaromatic compounds. 
Some subsets of these hydrocarbons exhibit different phases such as gas cluster, liquid, and solid phases, along with reactions. 
The universal hydrocarbon ab initio machine learning potential, referred to  
the RBCM potential in Eq. (\ref{eq:E_bcm}), was employed for its broader applicability.
The RBCM potential offered a significant advantage due to its faster scaling for training ($\mathcal{O}(nm^2/p^2)$), achieving comparable accuracy in predicting energy and forces to the local SGPR potential, referred to the sparse local experts (SLE) in Eq. (\ref{eq:E_alpha}) (Figure \ref{fig:test_errors}). 
Testing for gas-phase hydrocarbons covered alkane, alkene, isoalkane, alkadiene, alkatriene, cycloalkane, cycloalkane-methyl, cycloalkene, bicycloalkane, aromatic, and polyaromatic molecules (a total of 441 hydrocarbon molecules).
For gas cluster phase, we tested our RBCM potential with various ethene molecular clusters \ce{Et_n} ($n=2,3,...,25$).\cite{takeuchiTheoreticalInvestigationStructural2011}
For liquid hydrocarbons, local SGPR experts were trained with alkane (pentane, hexane, heptane, octane), alkene (pentene, hexene, octene), and aromatic (benzene, o-xylene, toluene) liquid phases. 
Solid phase testing covered various poly-aromatic solid ground state phases, such as benzene ($Pbca$), napthalene ($P121/c1$), antracene ($P 1 21/a 1$), tetracene ($P\bar{1}$), and pentacene crystals. 
Both local SGPR experts and the RBCM demonstrated energy prediction errors below the chemical accuracy. 
In general, the energy prediction of an individual SGPR expert were slightly better than the RBCM (shown in the Figure S1 $\sim$ S5). 
Regarding force predictions, RBCM matched or exceeded the accuracy of local SGPR experts. 
Notably, the RBCM model maintained vibrational frequency and zero-point energy predictions with an accuracy of less than 40 meV (320 cm$^{-1}$)

\subsection{Gas phase properties}

\subsubsection{Long-range $\pi$-$\pi$ interactions}

Hydrocarbons with nonpolar $\pi$-bonds,\cite{kimMolecularClustersPSystems2000} such as ethene and benzene, form $\pi$-complexes that are equilibrium structures, making them experimentally observable. These interactions play a pivotal role in various molecular phenomena, including the stacking of DNA base pairs, protein folding, molecular recognition, and properties in condensed phases. There's also an extensive body of literature exploring the interactions involving olefinic $\pi$-systems.\cite{kimMolecularClustersPSystems2000,tarakeshwarOlefinicVsAromatic2001,tsuzukiOriginAttractionDirectionality2002,sinnokrotHighAccuracyQuantumMechanical2006,leeUnderstandingAssemblyPhenomena2007,zhaoDensityFunctionalsNoncovalent2007,burnsComparingCounterpoiseCorrectedUncorrected2014}
Hence, it is essential to validate our hydrocarbon RBCM potential with the archetypal $\pi$-complexes: (i) olefinic ethene-ethene (ii) aromatic benzene-benzene, and (iii) olefinic-aromatic ethene-benzene interactions through nonpolar $\pi$-bonds, representing a challenging test-bed (Figure \ref{fig:pi-pi}). 

Each of these exemplary $\pi-\pi$ complexes poses a challenge due to their relatively low values of binding energies, which are on the order of several kcal/mol: 1.7 kcal/mol for ethene-ethene complex, (Fig. \ref{fig:pi-pi}a), 3.4 kcal/mol for benzene-benzene complex (Fig. \ref{fig:pi-pi}b), and 3.3 kcal/mol for ethene-benzene complex (Fig. \ref{fig:pi-pi}c). 
The errors in calculated binding energies are below a sub kcal/mol,  specifically 0.14, 0.4, and 0.2 kcal/mol for ethene-ethene, benzene-benzene, and ethene-benzene complexes, respectively. 
These minimal errors underscore the reliability of the RBCM potential in accurately representing nonpolar $\pi$-bond interactions within hydrocarbons.

\begin{figure*} [htp]
\includegraphics[width=17cm]{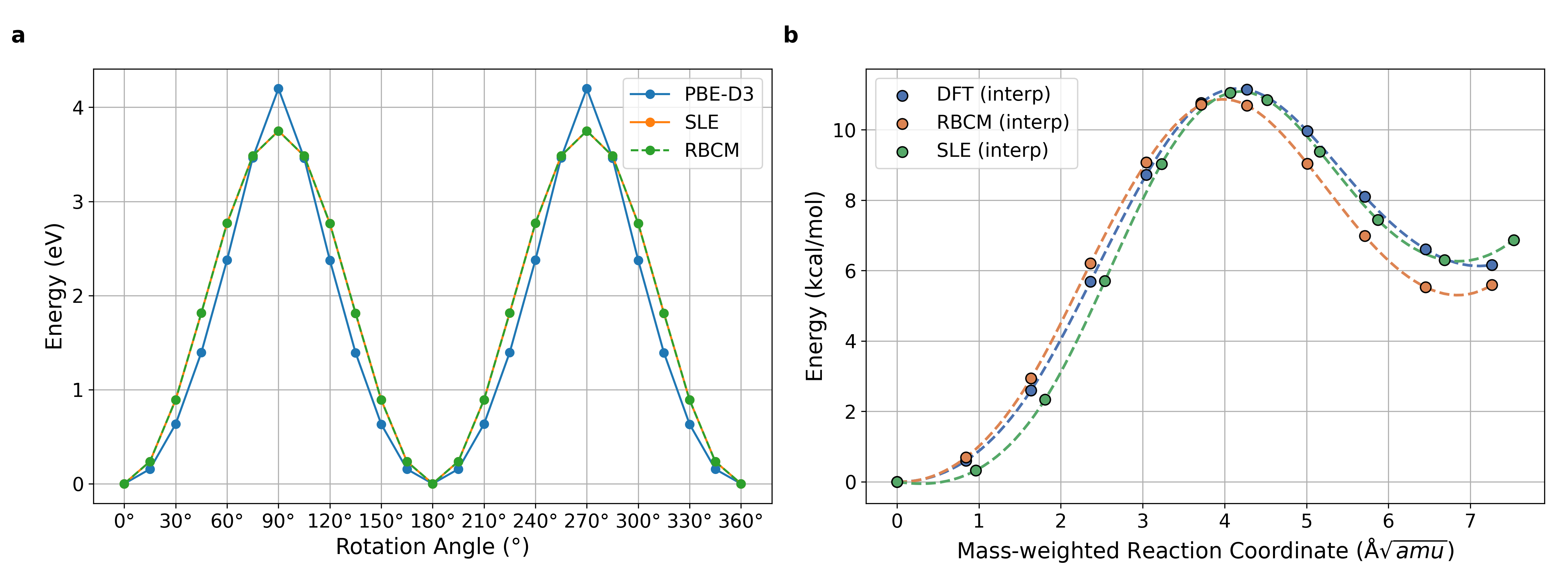}
\caption{Steric hindrance and puckering strain of hydrocarbons. 
(a) The energetic analysis (in eV) of rotational steric hindrance in spiropentane around the central carbon (Blue: PBE-D3, Orange: SLE, Green: RBCM)
(b) The interconversion pathway of puckering transitions in cyclohexane,
depicting the transformations among $^4C_1$, $^4H_3$, and $^1S_3$ puckered structures along the mass-weighted reaction coordinates (in \AA$\sqrt{amu}$).
}
\label{fig:steric-hindrance}
\end{figure*}

\subsubsection{Strains in hydrocarbons}
Steric effects play a pivotal role in chemistry by influencing the molecular geometry (conformational preferences, stability and consequent energy landscape). They also impact the rate of various reactions, such as polymerization, controlling active sites, as well as thermodynamic properties like density, heats of vaporization, boiling/melting points, and solubility.\cite{siuOptimizationOPLSAAForce2012,thomasConformationAlkanesGas2006,kangPuckeringTransitionsCyclohexane2018} 
Therefore, the incorporation of steric effects into a universal potential
greatly enhances the predictive accuracy of computational simulations for complex hydrocarbon systems.

To this end, we validate the RBCM potential with two prominent examples: transannular strain in the rotation motion of spiropentane (Fig. \ref{fig:steric-hindrance}a) and the puckering interchange conformational transitions of cyclohexane (Fig. \ref{fig:steric-hindrance}b).\cite{kangPuckeringTransitionsCyclohexane2018}
Spiropentane experiences substantial strain during rotation around the central carbon, with an activation energy of approximately 3.5 eV at the PBE-D3 level. 
The RBCM reproduces this energy landscape with an average energy error of 0.2 kcal/mol. 
It is well-known that the $^4H_1$ transition conformation appears when the ground state $^4C_1$ conformation converts into a $^1S_3$ conformation (Fig. \ref{fig:steric-hindrance}b).\cite{kangPuckeringTransitionsCyclohexane2018} 
Independent NEB calculations based on PBE-D3, SLE, and RBCM successfully identify the $^4H_1$ TS conformation. 

\subsection{Condensed phase properties}
\begin{figure*} [htp]
\includegraphics[width=17cm]{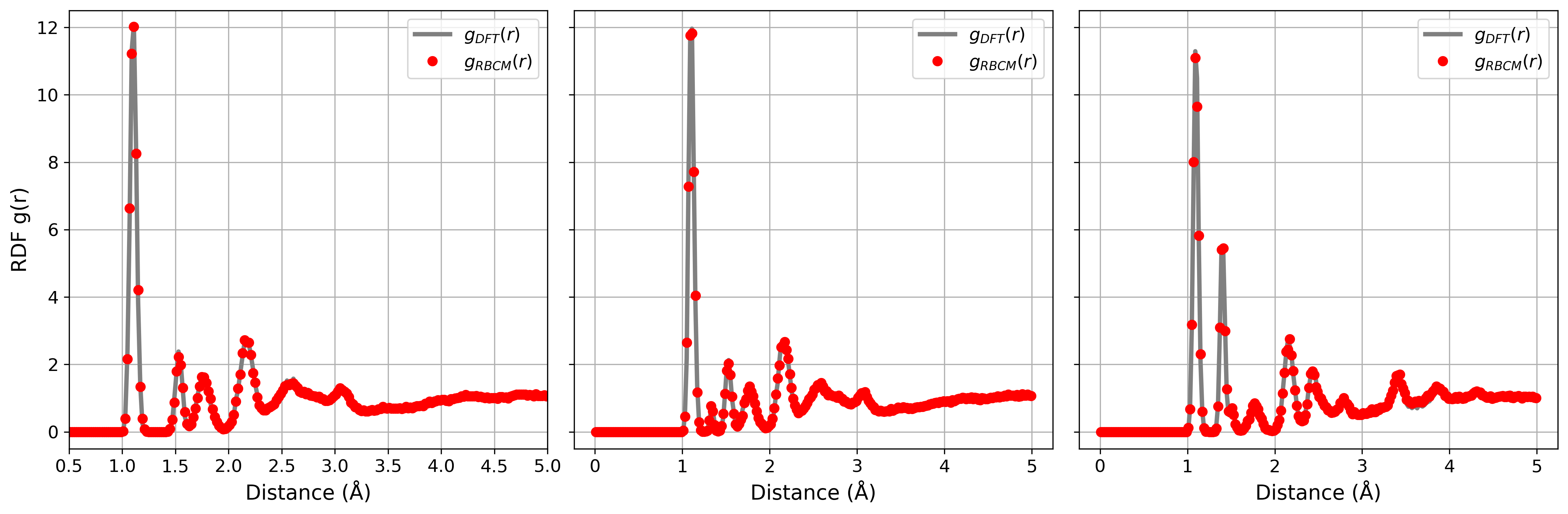}
\caption{Radial distributions $g(r)$ of 3 different hydrocarbons (a) octane liquid, (b) octene liquid, and (c) toluene liquid. The grey lines and red dots are DFT and RBCM calculations, respectively.}
\label{fig:RDF}
\end{figure*}

The reparameterization of torsional parameters in the OPLS-AA forcefields for long chain hydrocarbons has demonstrated a significant enhancement in achieving experimental accuracy for various thermodynamic properties of liquid hydrocarbons.
These thermodynamic properties include heats of vaporization, densities, and phase transition temperatures.\cite{thomasConformationAlkanesGas2006,siuOptimizationOPLSAAForce2012}
This implicates the need for careful consideration in parameterizing hydrocarbon forcefields, given the sensitivity of learning parameters for potential.
The challenge in accurately describing liquid phases with long hydrocarbon chains (with $n \ge 6$) arises from the complex torsional interactions of the carbon backbone.\cite{thomasConformationAlkanesGas2006,siuOptimizationOPLSAAForce2012}
Inaccuracies in the parameterization of torsional interactions for long chain hydrocarbons can lead to discrepancies in thermodynamic properties.
To validate the accuracy of the RBCM potential, we assessed its performance with alkanes \ce{C_nH_{2n+2}} and alkenes \ce{C_nH_{2n}} with different chain lengths ($n=5-8$) in liquid phases (Fig. \ref{fig:RDF} and Supplementary Fig. S4). 
Additionally, we tested the RBCM potential on aromatic liquids, including benzene, o-xylene, and toluene, to complete the validation for liquid phases.
Furthermore, we explored the RBCM potential's capability to accurately represent crystalline phases of various polyaromatic hydrocarbons, including benzene, naphthalene, anthracene, tetracene, and pentacene.

In Figure \ref{fig:RDF}, the RDFs obtained from MD simulations using the RBCM potential were compared with those obtained from MD simulations with DFT. 
Figure \ref{fig:RDF} show RDFs for octane liquid, octene liquid, and toluene liquid. 
These substances were chosen to represent alkanes and alkenes with long chain lengths,
as well as aromatic hydrocarbons. 
Our results demonstrate that the RBCM potential successfully reproduces RDFs for different hydrocarbons in both liquid and crystalline phases, indicating its reliability in capturing the structural properties of these systems. 
This underscores the potential utility of the universal RBCM in accurately describing the physical properties of various phases of both aliphatic and aromatic hydrocarbons.

\subsection{The Diels–Alder reaction}
Another application of RBCM hydrocarbon potential is the simulation of the reactive behavior of hydrocarbons. When the reaction center and transition states are unknown, the RBCM potential proves valuable for identifying them at an ab initio level. 
While previous empirical force fields could reasonably predict certain reactions,
they faced challenges, such as the failure to accurately simulate the 2+2 Diels-Alder reaction. 
These empirical force fields also fell short of achieving the numerical accuracy of ab initio simulation.\cite{vanduinReaxFFReactiveForce2001} 
The Diels-Alder reaction stands as a cornerstone in synthetic organic chemistry, offering an elegant and efficient pathway to form six-membered rings. 
However, due to its multi-reference nature and various reaction channels (stepwise, concerted, etc.), the complete understanding of the Diels-Alder mechanism remains elusive.
Learning the Diels–Alder reaction at an ab initio level can offer crucial insights into its mechanism and unveil strategies to optimize its efficacy for modern synthetic applications. 
To this end, we demonstrate a reaction pathway search of the cycloaddition of 1,3-butadiene and ethene. The accuracy of the RBCM potential hints at its potential as an ab initio universal potential for a wide range of organic chemistry compounds.\cite{vanduinReaxFFReactiveForce2001}

\begin{figure} [htp]
\includegraphics[width=8cm]{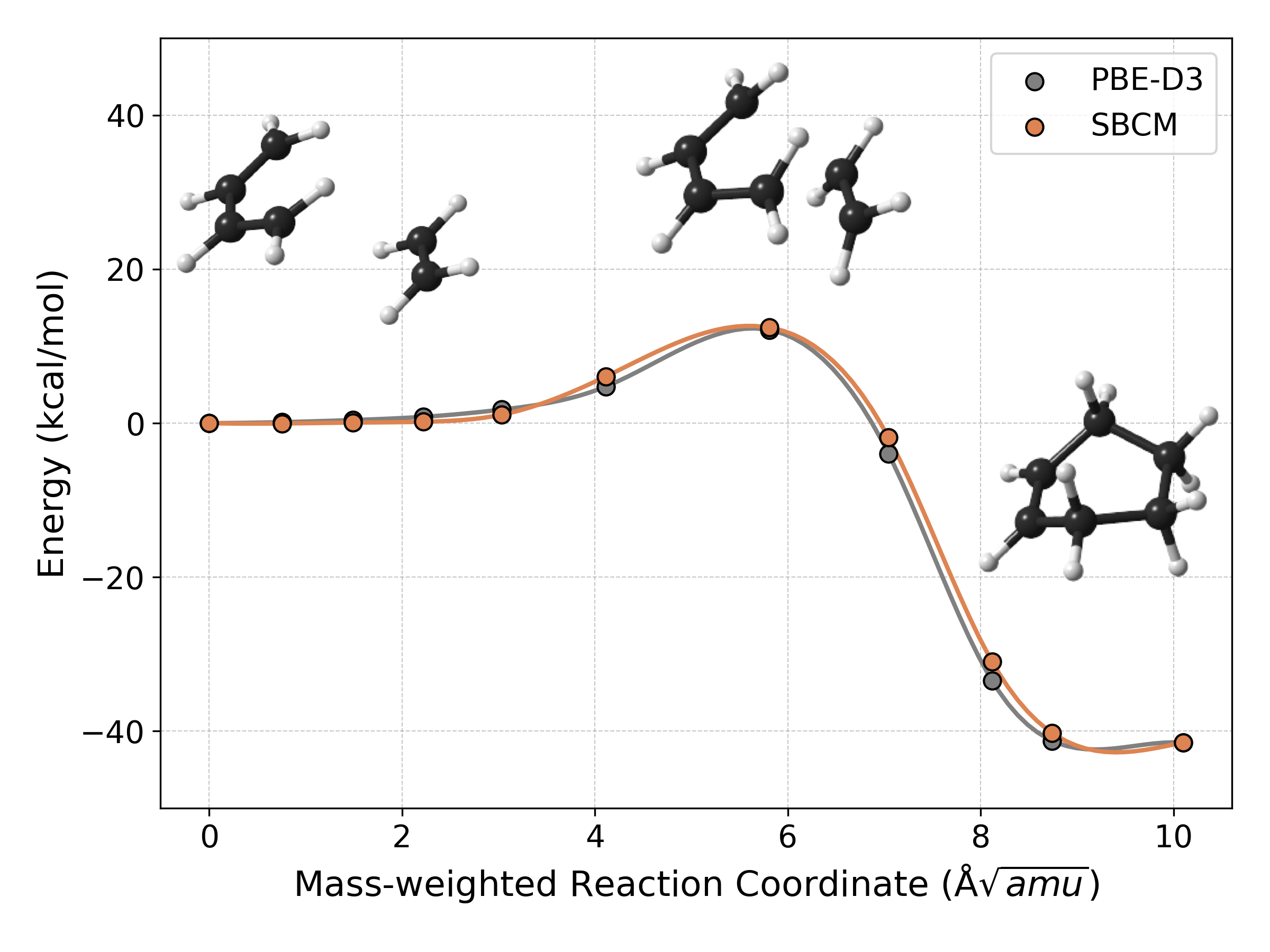}
\caption{Potential energy profile (in kcal/mol) of PBE-D3 (gray line) and SBCM potential (orange line) for the Diels-Alder reaction, cycloaddition between 1,3-butadiene and ethene, along the mass-weighted reaction coordinates (in \AA$\sqrt{amu}$). The depicted pathway elucidates the concerted mechanism inherent to this archetypal reaction, showcasing the synchronous formation of new sigma bonds that lead to the cyclohexene product.}
\label{fig:04-DA-NEB}
\end{figure}

Accurate estimation of the energy barriers at transition states and the relative energy differences between the initial and product states is of paramount importance in understanding reaction kinetics and mechanisms. 
The pathway of the Diels–Alder reaction involving 1,3-butadiene and ethene was estimated by the PBE-D3 and the RBCM potential, as depicted in Fig. \ref{fig:04-DA-NEB}. 
The obtained results revealed tiny errors in the reaction barrier (0.31 kcal/mol) and  the relative energy difference (0.01 kcal/mol).
These findings underscore the accuracy and transferrability of the RBCM potential in faithfully representing  energy landscapes associated with the Diels-Alder reaction.
The minimal discrepancies observed in the calculated values further highlight the robustness of the RBCM potential in accurately capturing the intricate details of the reaction energetics. 
Overall, these outcomes contribute to the confidence in the reliability of the RBCM potential for modeling and understanding the Diels–Alder reaction, providing valuable insights into its kinetics and underlying mechanisms.


\section{Discussion}

The findings regarding the gas phase properties of various hydrocarbons present a significant advancement in our understanding of hydrocarbon interactions, particularly through the lens of the BCM potential. The BCM potential's ability to accurately predict long-range $\pi$-$\pi$ interactions in hydrocarbons is a breakthrough, especially given its efficiency and the challenging nature of these interactions. These $\pi$-$\pi$ interactions are crucial in many biological and chemical processes, such as the stacking of DNA base pairs and molecular recognition. The low error margins in binding energies of these complexes reinforce the potential's reliability.

The results also highlight the significance of steric strains in hydrocarbons, which are essential in determining molecular geometry and reaction rates. The RBCM potential’s capability to model transannular strain and puckering transitions in molecules like spiropentane and cyclohexane is an invaluable tool for understanding these complex phenomena.

In the realm of condensed phase properties, the RBCM potential showcases its versatility and accuracy. The challenges posed by long-chain hydrocarbons in liquid phases, due to the intricate dihedral interactions, are well-addressed by the potential. Its proficiency in reproducing the radial distribution functions of various hydrocarbons, as illustrated in Figure \ref{fig:RDF}, is commendable. This aspect is critical, as the accurate modeling of these properties is essential for understanding the thermodynamics of hydrocarbons and designing better materials and chemicals.

The application of the RBCM potential in simulating the Diels–Alder reaction is particularly noteworthy. This reaction is a cornerstone in synthetic organic chemistry, and the potential’s ability to accurately simulate its mechanism opens new avenues in synthetic applications. The reaction pathway search for the cycloaddition of 1,3-butadiene and ethene, as shown in Figure \ref{fig:04-DA-NEB}, is a testament to the potential's capability to operate at an ab initio level. The minimal errors in estimating the energy barrier and the relative energetics further solidify its role in understanding complex organic reactions.

These findings have far-reaching implications. The RBCM potential, with its accuracy and efficiency, represents a significant stride in computational chemistry, especially in the modeling of complex hydrocarbon systems and reaction mechanisms. Its ability to handle different phases and reactions with high precision is likely to be a game-changer in the field.

Future research should focus on expanding the application scope of the RBCM potential to include a broader range of chemical reactions and compounds. This could further validate its versatility and contribute to a deeper understanding of chemical processes at the molecular level. Additionally, integrating the RBCM potential into computational tools used in industry and academia could revolutionize the way chemists and material scientists design experiments and synthesize new materials.

In summary, the RBCM potential's demonstration of robustness across various phases and reactions of hydrocarbons presents a significant advancement in computational chemistry. Its implications for future research and practical applications are profound, paving the way for more accurate and efficient modeling of complex chemical systems.

\section{Methods}
\subsection{Computational details for ML potential}
To obtain a machine learning potential $E_\alpha$ of each expert model, we performed the canonical ensemble (NVT) molecular dynamics simulations using a Nose-Hoover thermodstat and Parrinello-Rahman dynamics, implemented in the atomic simulation environment (ASE) package. The MD simulations ran for 3$\sim$10 ps at 300 K with a 0.5 fs time step. 
The reference potential energy surfaces are from density functional theory (DFT) calculations, performed using the Vienna ab initio simulation package (VASP)~\cite{kresseEfficiencyAbinitioTotal1996,PhysRevB.54.11169} with Perdew-Burke-Ernzerhof (PBE) functionals\cite{perdewRationaleMixingExact1996} and van der Waals corrections (D3).\cite{grimmeConsistentAccurateInitio2010}
We employed projector augmented wave\cite{blochlProjectorAugmentedwaveMethod1994} pseudopotentials with a 400 eV energy cut-off, setting the convergence criterion for the electronic energy difference at 10$^{-4}$ eV.
The molecules were placed in  the center of a cubic cell with a 15 \AA ~vacuum interlayer distance, and the Brillouin zone was sampled using the $\Gamma-$point. 

The local SGPR potential\cite{hajibabaeiSparseGaussianProcess2021} was trained using AUTOFORCE package.\cite{AutoForce} 
We examined expert models tailored for linear or branched systems (C1-C8 alkane, C3-C10 isoalkane, C2-C6 alkene, C3-C6 brached alkene, C3-C8 alkadiene, and C7-C9 alkatriene), cyclic systems (C3-C10 cycloalkane, C3-C8 cycloalkene, and C4-C10 methyl-cycloalkane, bicycloalkane), and aromatic systems (bezene, toluene, xylene, trimethylbenzene, and mesitylene) in constructing the universal model. 
Additionally, to account for intermolecular interactions, expert models were trained for the liquid phases of alkanes (methane, ethane, propane, butane, pentane, hexane, heptane, octane, and decane), alkene (ethylene), and aromatic (benzene), as well as for the solid phases (benzene, napthalene, anthracene, tetracene, and pentacene).


\section{Acknowledgements}
The authors are grateful for computational resources provided by the Korea Institute of Science and Technology Information (KISTI) for the Nurion cluster (KSC-2021-CRE-0542, KSC-2022-CRE-0115). C.W.M. acknowledges the support from the National Research Foundation of Korea (NRF) grant funded by the Korea government (MSIT) (No. NRF-2022R1C1C1010605).
SYW and CWM acknowledge the support from the National Research Foundation of Korea (NRF) grant funded by the Korea government (MSIT) (No. RS-2023-00222245).

\bibliographystyle{apsrev4-1}
\bibliography{hydrocarbons}
\end{document}


\title{A Sparse Bayesian Committee Machine Potential for Hydrocarbons}

\author{Soohaeng Yoo Willow}
\affiliation{Department of Energy Science, Sungkyunkwan University, Seobu-ro 2066, Suwon, 16419, Korea}

\author{Gyung Su Kim}
\affiliation{Department of Energy Science, Sungkyunkwan University, Seobu-ro 2066, Suwon, 16419, Korea}

\author{Miran Ha}
\affiliation{Department of Energy Science, Sungkyunkwan University, Seobu-ro 2066, Suwon, 16419, Korea}

\author{Amir Hajibabaei}
\affiliation{Department of Energy Science, University of Cambridge, UK}

\author{Chang Woo Myung}
\email{cwmyung@skku.edu}
\affiliation{Department of Energy Science, Sungkyunkwan University, Seobu-ro 2066, Suwon, 16419, Korea}

\date{\today}

\maketitle
\tableofcontents
\newpage




\section{Gas phase hydrocarbons}
\subsection{Test errors for gas phase hydrocarbons}

The following hydrocarbon groups were considered for the gas phase:
\begin{itemize}
    \item alkane : methane, ethane, propane, butane, pentane, hexane, heptane, octane
    \item alkadiene: butadiene, pentadiene, 1,4-hexadiene, 1,5-hexadiene, 2,4-hexadiene, heptadiene, and octadiene. 
    \item alkatriene: heptatriene, octatriene, nonatriene
    \item isoalkane (branched): isobutane, isopentane, isohexane, isoheptane, isooctane, isononane, isodecane
    \item alkene: ethylene, propylene, butene, pentene, hexene, octene, isobutylene, isopentene, isohexene, isoheptene
    
    \item alkene-23 (C=C at second or third): 2-butene, cis-2-butene, 2-pentene, cis-2-pentene, trans-2-pentene, 2-hexene, 3-hexene, cis-3-hexene, trans-3-hexene
    
    \item branched-alkene: isobutylene, 2,3-dimethyl-1-butene, 2-ethyl-1-butene, 2-methyl-1-butene, cis-3-methyl-2-pentene, trans-3-methyl-2-pentene
    
    \item aromatic: benzene, mesitylene, m-xylene, O-xylene, toluene, 1,2,3-trimethyl-benzene, 1,2,4-trimethyl-benzene
    
    \item cycloalkane: cyclopropane, cyclobutane, cyclopentane, cyclohexane, cycloheptane, cyclooctane, cycleononane, cyclodecane
    
    \item cycloalkane-methyl: 1,1-dimethyl-cyclopropane, 1,1-dimethyl-cyclobutane, 1,1-dimethyl-cyclopentane, 1,1-dimethyl-cyclohexane, 
    1,2-dimethyl-cyclopropane, 1,2-dimethyl-cyclobutane, 1,2-dimethyl-cyclopentane, 1,2-dimethyl-cyclohexane,
    1,2-dimethyl-cyclooctane, 
    1,3-dimethyl-cyclobutane, 1,3-dimethyl-cyclopentane, 1,3-dimethyl-cyclohexane, 
    1,4-dimethyl-cyclooctane, 1,5-dimethyl-cyclooctane,
    methyl-cyclopropane, methyl-cyclobutane, methyl-cyclopentane, methyl-cyclohexane, methyl-cycloheptane, methyl-cyclooctane

    \item cycloalkene: cyclopropene, cyclobutene, cyclopentene, cyclohexene, cycloheptene, cyclooctene, trans-cyclooctene

    \item bicycloalkane: 1-cyclopropyl-1-methylcyclohexane, cyclobutane, bicycloheptane, bicyclononane, bicyclodecane, 
    methylbiocyclo-nonane, spiropentane, spirodecane

    \item pah: anthracene, benzo[a]pyrene, benzo[b]fluoranthene, benzo[ghi]perylene, benzo[j]fluoranthene, benzo[k]fluorathene, chrysene, coronene, dibenzo[a,e]pyrene, dibenzo[a,h]anthracene, dibenzo[a,i]pyrene, dibenzo[a,l]pyrene, fluoranthene, hexacene, heptacene, naphthalene, ovalene, pentacene, phenanthrene, picene, pyrene, tetracene, triphenylene

\end{itemize}

\begin{figure}[h!]
    \centering
    \includegraphics[width=1.0\textwidth]{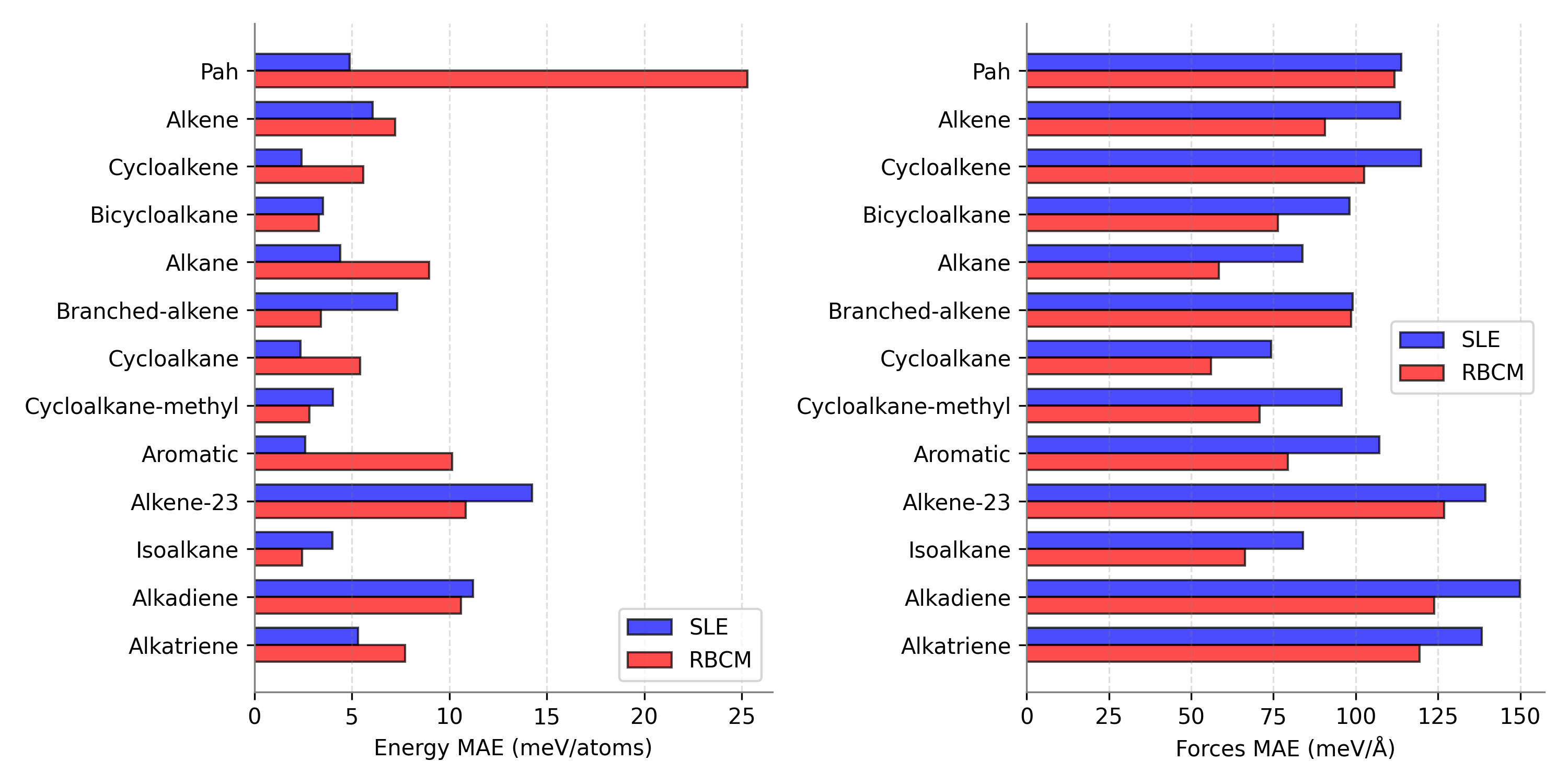}
    \caption{Comparative analysis of the mean absolute error (MAE) in energy and forces across various hydrocarbons. The left panel illustrates the energy MAE per atom (in meV/atoms) while the right panel shows the forces MAE (in meV/Å). Two computational methods are evaluated: SLE (blue bars) and RBCM (red bars).}
    \label{fig:gas-phase_geometries}
\end{figure}

\newpage 
\subsection{Test errors for gas phase ethene clusters}
The ethene clusters, Et$_n$ (n = 2 $\sim$ 25), in the gas phase were considered.

\begin{figure}[h!]
    \centering
    \includegraphics[width=1.0\textwidth]{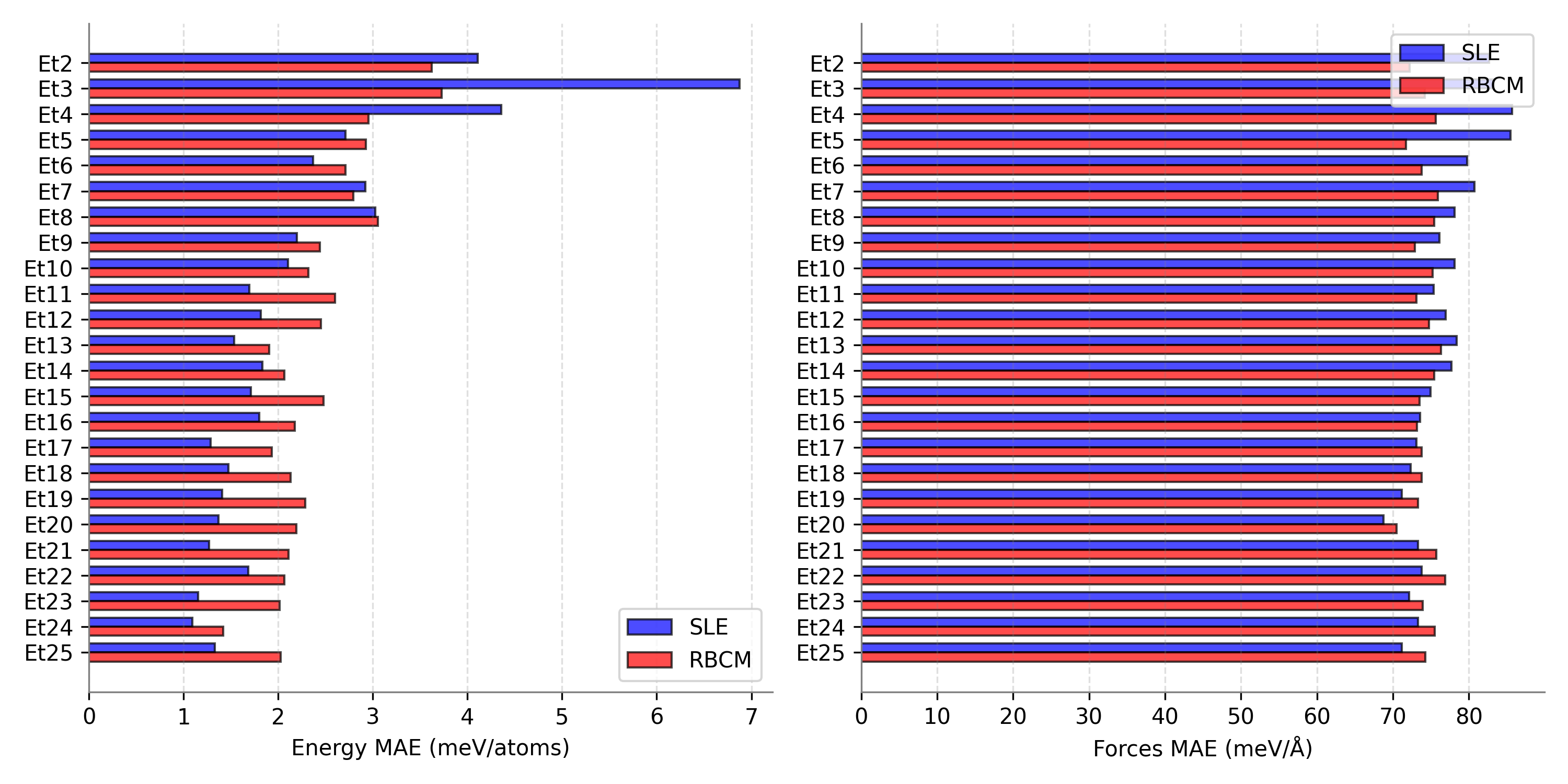}
    \caption{Comparative analysis of the mean absolute error (MAE) in energy and forces of ethene (Et) molecular clusters \ce{Et_n} ($n = 2, 3, ..., 25$). The left panel illustrates the energy MAE per atom (in meV/atoms) while the right panel shows the forces MAE (in meV/Å). Two computational methods are evaluated: SLE (blue bars) and RBCM (red bars).}
    \label{fig:gas-phase_geometries}
\end{figure}

\newpage
\subsection{Test errors for vibrational frequency of hydrocarbons}

We measure the zero-point energy normalized MAE excluding 3 translational and 3 rotational modes:
$$\frac{1}{3N_{atom}-6}\sum^{3N_{atom}-6}_i \frac{|\nu_i^{DFT}-\nu_i^{ML}|}{\nu_{ZPE}}$$. 

\begin{figure}[h!]
    \centering
    \includegraphics[width=1.0\textwidth]{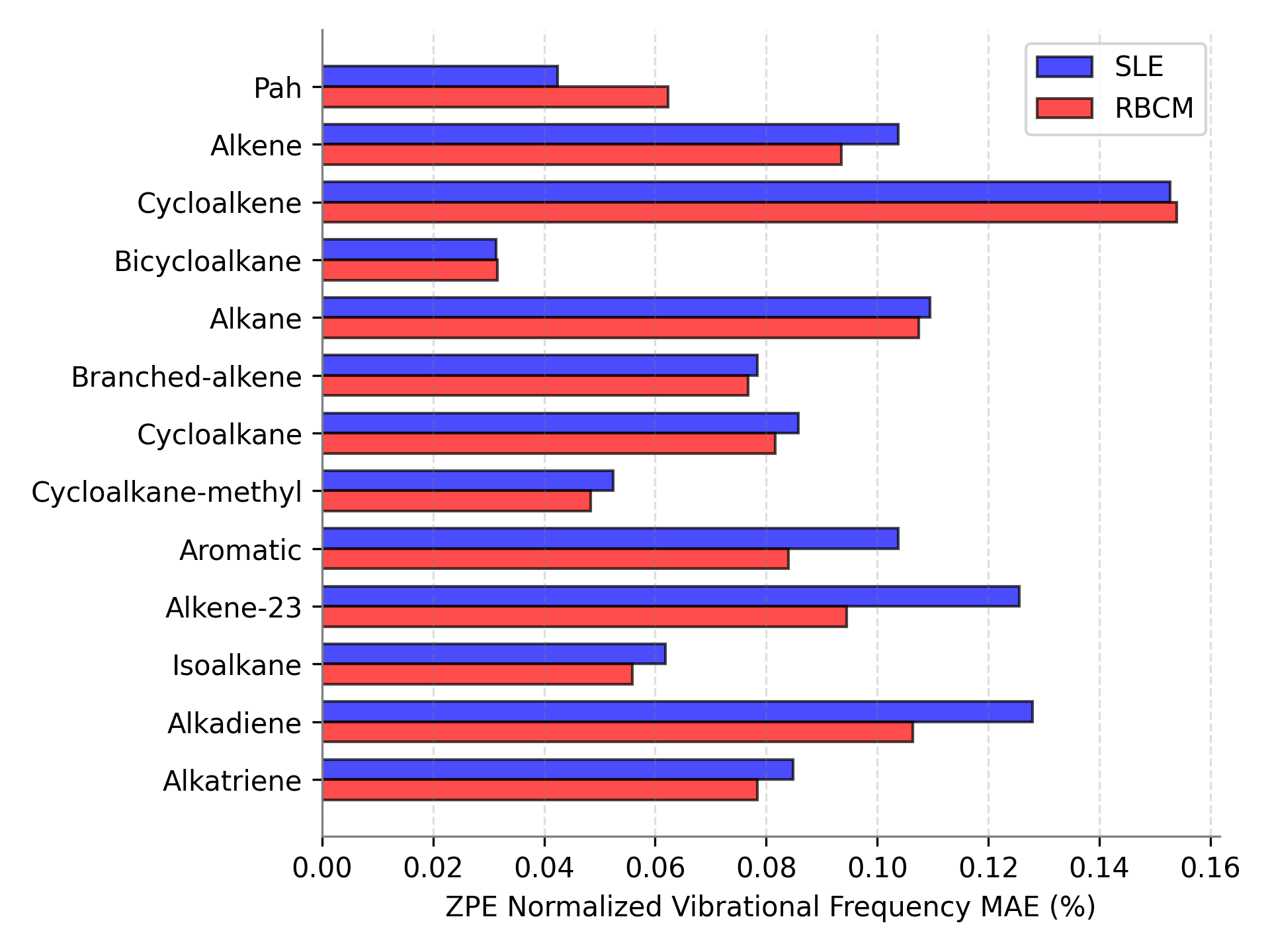}
    \caption{Comparative analysis of the zero-point energy normalized MAE of vibrational frequency of various hydrocarbons. Two computational methods are evaluated: SLE (blue bars) and RBCM (red bars).}
    \label{fig:gas-phase_geometries}
\end{figure}

\newpage 
\section{Liquid phase hydrocarbons}
\subsection{Test errors for liquid hydrocarbons}
The following hydrocarbon groups were considered for the liquid phase:
\begin{itemize}
\item alkane-liq: pentane, hexane, heptane, octane
\item alkene-liq: hexene, heptene, octene
\item aromatic-liq: benzene, toluene, o-xylene
\end{itemize}

To train SGPR models for the liquid phase, NVT MD simulations were performed for a duration of 10 ps at a temperature of 300 K.

\begin{figure}[h!]
    \centering
    \includegraphics[width=1.0\textwidth]{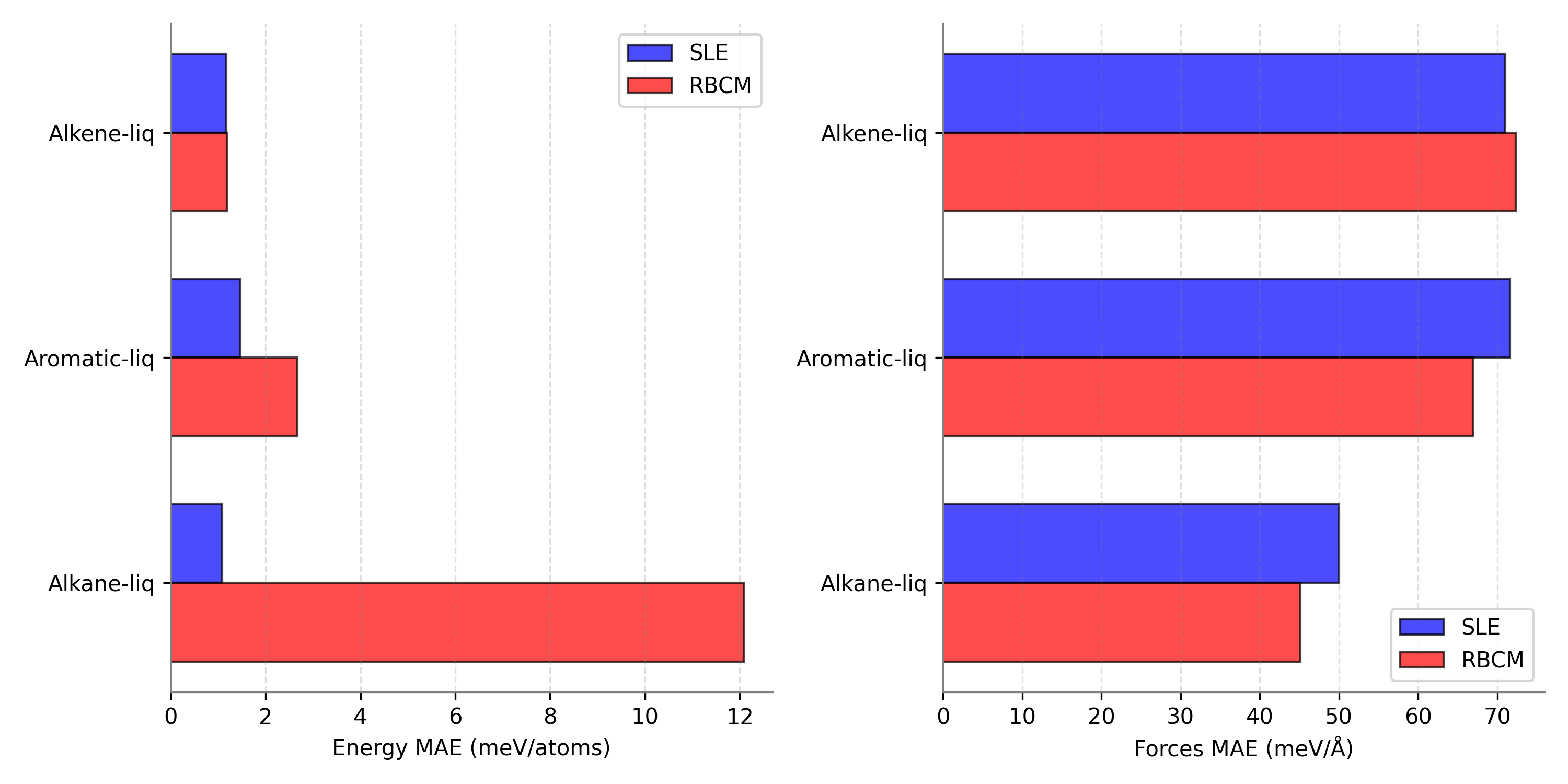}
    \caption{Comparative analysis of MAE of various liquid hydrocarbons. Two computational methods are evaluated: SLE (blue bars) and RBCM (red bars).}
    \label{fig:gas-phase_geometries}
\end{figure}

\newpage
\section{Solid phase hydrocarbons}
\subsection{Test errors for solid hydrocarbons}
The following PAH hydrocarbons were considered for the solid phase:
\begin{itemize}
\item PAH: tetracene, napthalene, anthracene, pentacene, benzene
\end{itemize}

\begin{figure}[h!]
    \centering
    \includegraphics[width=1.0\textwidth]{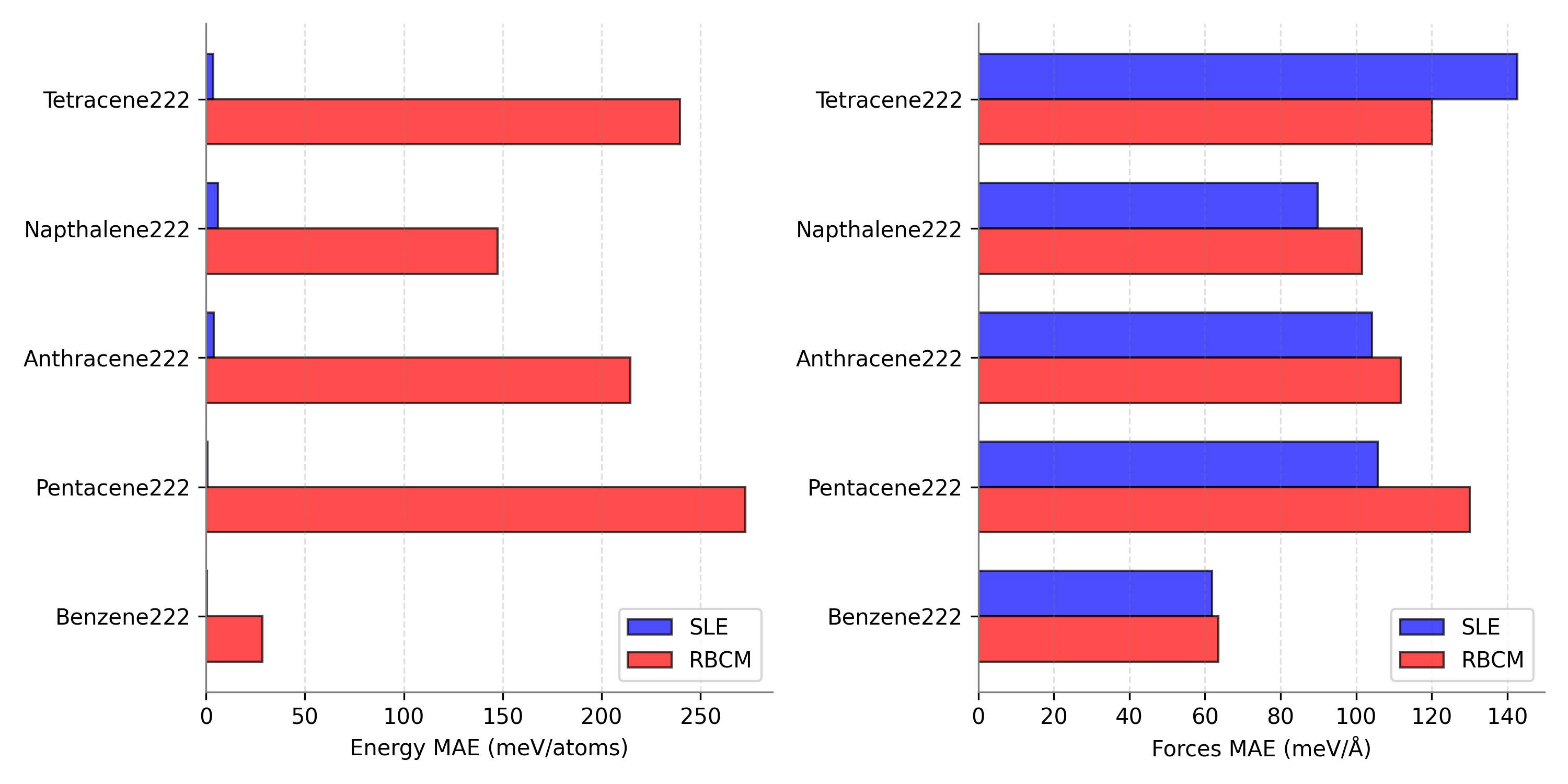}
    \caption{Comparative analysis of MAE of various solid PAH hydrocarbons. Two computational methods are evaluated: SLE (blue bars) and RBCM (red bars).}
    \label{fig:gas-phase_geometries}
\end{figure}